\documentclass{article}
\usepackage[dblblindworkshop,final]{neurips_2026}
\workshoptitle{Personalized, Aligned, Long-Term Memory for AI Systems (PALM)}
\usepackage[utf8]{inputenc}
\usepackage[T1]{fontenc}
\usepackage{hyperref}
\usepackage{url}
\usepackage{xcolor}
\usepackage{graphicx}
\usepackage{booktabs}
\usepackage{array}
\usepackage{amsfonts}
\usepackage{nicefrac}
\usepackage{microtype}
\usepackage{amsmath}
\usepackage{placeins}
\usepackage{needspace}
\usepackage{flafter}
\usepackage{tikz}
\usepackage{fvextra}
\DefineVerbatimEnvironment{Prompt}{Verbatim}{fontsize=\footnotesize,breaklines=true}
\usetikzlibrary{positioning,arrows.meta}
\hypersetup{hidelinks,pdfauthor={Jeffrey Chu, Anjali Joseph, Rivan Mandot, Feker Demere, Vincent Li},pdftitle={Peer Memory Changes Who LLM Agents Consult Without a Detected Gain in Answer Accuracy}}

\title{Peer Memory Changes Who LLM Agents Consult Without a Detected Gain in Answer Accuracy}
\author{%
  Jeffrey Chu \\
  New York Institute of Technology
  \And
  Anjali Joseph \\
  Georgia State University
  \And
  Rivan Mandot \\
  Algoverse AI Research
  \AND
  Feker Demere \\
  Algoverse AI Research
  \And
  Vincent Li \\
  Columbia University
}

\begin{document}
\maketitle

\begin{abstract}
Multi-agent LLM systems often reuse the same agents across tasks, creating an opportunity to draw on past interactions when deciding whom to consult. We study this form of memory with PeerMem, which saves MACE's LinUCB state and reuses it when the same agents return. Using six agents, a fixed set of 80 HotpotQA questions, and 12 paired lineages per model, we compare cold, warm, and shuffled initialization in an answer-exchange protocol with reference-answer feedback. Under stable evidence access, warm memory lowers pooled hindsight regret and increases consultation of other evidence holders on Qwen3-32B and Gemma-4-31B-it, while final token $F_1$ differences remain uncertain. Qwen interaction traces further show that warm agents more often encounter peer answers that score above their own prior answers. We extend the evaluation to Qwen3.5-9B using freshly learned memory to compare warm and cold initialization under reassigned and paired stable roles. The primary changed-minus-stable final-$F_1$ interaction is $-1.38$ points, with a 95\% interval of $[-3.50,+0.73]$. Together, these experiments trace the effects of peer memory through consultation, answer revision, and final task performance.
\end{abstract}

\section{Introduction}

Multi-agent large language model (LLM) systems often reuse the same agents across tasks, allowing past interactions to inform later decisions about whom to consult \citep{wu2023autogen,li2023camel,guo2024survey}. Remembering which peers were useful can reduce repeated exploration, provided that their past performance remains informative. The value of this experience also depends on what a selected peer communicates and how the reader uses that information to revise its answer.

We study this problem in the HotpotQA setting used by MACE, where six agents each receive one passage: two receive evidence and four receive distractors, with these roles hidden from the agents \citep{mace2026,hotpotqa2018}. In the stable-role experiment, the same IDs retain their roles across two sets of 40 questions within a paired replicate, which we call a lineage. This makes past experience relevant to later consultation, since peer usefulness primarily reflects stable access to evidence and the selector receives reference-answer feedback after each revision. Our extension, \textbf{PeerMem}, saves the LinUCB statistics learned for each selector-peer pair and reuses them when the same IDs return \citep{li2010linucb}. On the second set of questions, we compare empty state, correctly matched saved state, and the same state shuffled across peer IDs.

Our contribution is a controlled empirical study of how persistent peer memory shapes collaboration. The cold, warm, and shuffled conditions measure changes in consultation under stable evidence access, while interaction traces connect the quality of consulted answers to the revisions that follow. We then move evidence to former distractor holders in a further study, pairing the reassigned conditions with stable controls to examine memory under changing information access. This design evaluates peer memory across three stages of collaboration: choosing whom to consult, revising an answer, and completing the task.

\section{Related Work}

\subsection{Multi-agent coordination and peer selection}

Multi-agent LLM systems support collaborative reasoning by exchanging answers, critiques, and intermediate results \citep{wu2023autogen,li2023camel,du2023debate}. Effective collaboration depends on both coordination and the use of the information received: debate does not consistently improve over simpler prompting \citep{smit2024mad}, and communication or verification failures can undermine its benefits \citep{cemri2025fail}. These findings motivate evaluating peer selection together with the answers agents produce after consultation.

MACE learns pair-specific LinUCB statistics for peer selection; its July 2026 version evaluates reuse with frozen parameters on a later task block and transfer from HotpotQA to 2WikiMultihopQA \citep{mace2026,li2010linucb}. PeerMem builds on this selector by comparing cold, correctly indexed, and shuffled initialization while learning continues online. Across stable and reassigned evidence access, we track how this initialization affects both consultation and subsequent answer revision.

\subsection{Persistent memory, partner models, and reputation}

Agent memory retains reflections, skills, and workflows for later tasks \citep{shinn2023reflexion,zhao2023expel,wang2024awm}. G-Memory organizes collaboration experience into a hierarchy of insight, query, and interaction graphs that agents can reuse across trials \citep{zhang2025gmemory}. Explicit Trait Inference models partners' warmth and competence from interaction histories, reuses frozen profiles in later interactions, and tests profile transfer between Stag Hunt and Prisoner's Dilemma \citep{abdurahman2026eti}.

Several methods focus on peer reliability. $\Sigma$-Mem updates peer-competence and peer-relationship memories from correctness feedback and evaluates counterfactual reliability shifts, with a direct routing readout that returns the selected peer's answer \citep{feng2026sigmamem}. Our protocol follows the additional step in which the reader revises its own answer after consultation. Related approaches include DRF, which combines an interactive rating network, reputation scores, and UCB-based agent selection \citep{lou2026drf}, and CADMAS-CTX, which maintains context-conditioned capability posteriors for uncertainty-aware delegation and evaluates warm starts followed by task-mixture shifts \citep{cadmasctx2026}. Together, these studies motivate examining how a selector's accumulated experience affects whom agents consult and the answers they subsequently produce.

This question also builds on classical reputation systems and bandits that learn from historical observations \citep{josang2002beta,teacy2006travos,shivaswamy2012history,zhang2019warmstart}. By holding the selector fixed and varying its initialization, we study the value of past experience in a setting where access to evidence is the main source of differences in peer usefulness.

\section{Method}
\label{sec:method}

\subsection{Task and interaction protocol}

We use the HotpotQA distractor setting, in which each question contains two evidence passages and eight distractors \citep{hotpotqa2018}. For each question, we distribute the two evidence passages and four sampled distractors among six agents, giving each agent one passage. The stable-role experiment keeps evidence with the same two IDs throughout both task blocks, so the passage contents change across questions while the hidden roles remain fixed.

Each question begins at revision round $t=0$, when every agent answers from its own passage, and continues through three rounds of consultation and revision. In round $t$, each agent selects an ID, reads that peer's answer from round $t-1$, and revises its own answer; all agents complete a round before the next begins. The exchange contains only answers, while each agent's passage remains private and evidence-role labels remain hidden.

\begin{figure}[t]
    \centering
    \begin{tikzpicture}[
  font=\footnotesize,
  box/.style={draw=black!55, rounded corners=2pt, align=center,
    text width=3.65cm, inner sep=5pt, minimum height=1.65cm},
  >={Stealth[length=5pt]}
]
\node[box,fill=black!3] (learn) {\textbf{Layer 1: 40 questions}\\[3pt]
  Six agents; two evidence\\ holders (roles hidden)\\[3pt]
  MACE learns pair-indexed state};
\node[box,fill=blue!4,right=0.42cm of learn] (memory) {\textbf{Layer 2 initialization}\\[3pt]
  Cold: empty state\\
  Warm: original peer IDs\\
  Shuffled: permute non-self IDs};
\node[box,fill=black!3,right=0.42cm of memory] (evaluate) {\textbf{Layer 2: 40 new questions}\\[3pt]
  Same hidden evidence roles\\
  Continue online updates\\[3pt]
  Measure consultation and $F_1$};
\draw[->] (learn) -- (memory);
\draw[->] (memory) -- (evaluate);
\node[draw=black!40,align=center,inner sep=4pt,text width=12.3cm,
  below=0.36cm of memory] (loop) {\textbf{Each revision:}
  choose among all six IDs $\longrightarrow$ read the prior extracted answer
  $\longrightarrow$ revise $\longrightarrow$ update from reference-answer $F_1$};
\end{tikzpicture}
    \caption{Stable-role protocol with evidence assigned to the same two IDs across task blocks and roles hidden from the agents. Each question consists of an initial answer and three synchronous revisions per agent. The MACE conditions differ in initialization and all update online using reference-answer rewards; random selection provides an additional baseline.}
    \label{fig:workflow}
\end{figure}
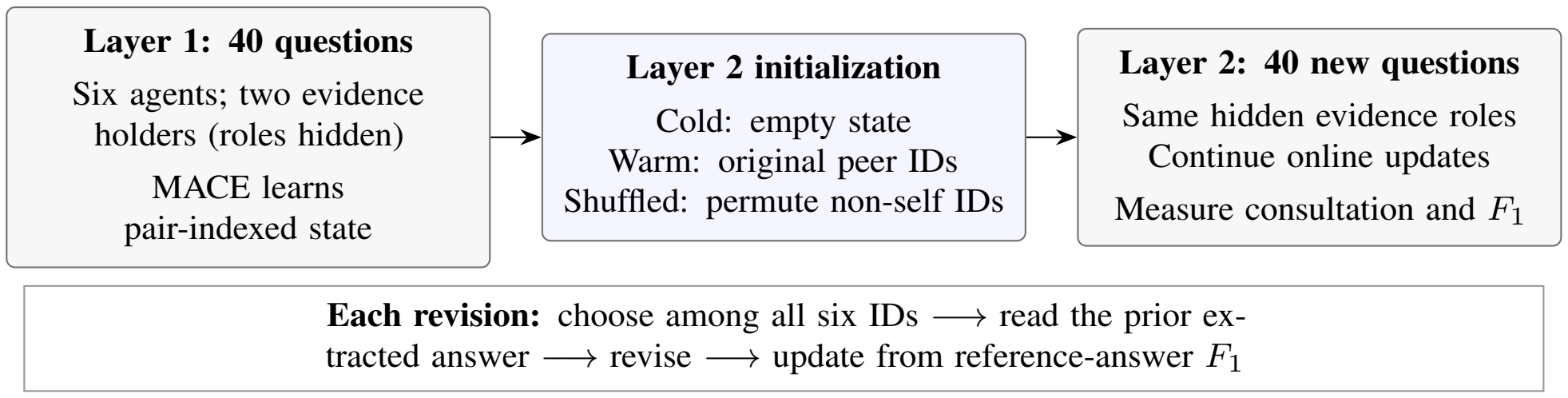

MACE, UCB1, in-context, and random selection can choose any of the six IDs, including self, in which case the agent receives its own previous answer. Uniform random selection therefore chooses self with probability $1/6$, and the fixed-cycle baseline always selects the next ID. Figure~\ref{fig:workflow} summarizes this protocol and the memory intervention between task blocks.

\subsection{Peer selection and persistent memory}
\label{sec:peer_memory}

MACE represents each possible read with a normalized nine-feature context vector computed from the agents' responses and interaction history \citep{mace2026}. The vector combines three $n$-gram diversity features, three response-graph centrality features, the selector-peer pair's past mean reward, the normalized revision round, and a bias term. Appendix~\ref{app:reproducibility} gives the full definition.

For each ordered pair of selecting agent $i$ and candidate peer $j$, LinUCB maintains a design matrix $A_{ij}$ and reward vector $b_{ij}$, along with the pair's selection count and mean reward \citep{li2010linucb}. MACE selects the candidate with the largest predicted reward plus an uncertainty bonus, then updates that pair's statistics using the reward from the reader's revised answer. PeerMem saves these pair-indexed statistics so that they can initialize a later run.

We vary this initialization through three conditions: a cold start begins with empty state, a warm start loads the saved state under the original peer IDs, and a shuffled start assigns each of the five non-self records to a different peer ID while leaving the self record in place. All three conditions continue updating their state after initialization, allowing us to compare their behavior under the same feedback rule.

\subsection{Reward and evaluation measures}
\label{sec:metrics}

We evaluate answers using exact match (EM) and bag-of-token $F_1$ under the SQuAD convention, which normalizes text by lowercasing, removing punctuation and articles, and collapsing whitespace \citep{rajpurkar2016squad}. Final accuracy averages the six agents' individual scores after revision round $T=3$, giving 240 responses from 40 questions for each lineage and condition. Comparisons then use the 12 paired lineage means.

Let $s_{q,i,t}$ be the selecting agent's token $F_1$ after revision round $t$. We reward a selection using the reader's resulting answer:

\begin{equation}
r_{q,i,t}=\tfrac{1}{2}\left[(s_{q,i,t}-s_{q,i,t-1})+s_{q,i,t}\right].
\label{eq:reward}
\end{equation}

The reward lies in $[-0.5,1]$ and credits both improvement and retention of a high-scoring answer. In both layers, the evaluator computes it from the reference answer and uses it to update the selector, while keeping the reference answer hidden from the answering agents.

\paragraph{Evidence-holder reads.}
We identify evidence holders from the seeded passage assignment and measure consultation in two ways. The first is the fraction of all selections directed to either evidence-holder ID, including self-selection, and has a uniform expectation of $2/6=1/3$; this is the measure shown in Figure~\ref{fig:selection_trajectory}. The external evidence-holder read rate counts only selections of an evidence holder other than the reader, using all selections as its denominator. With a uniform expectation of $5/18$, this rate measures how often the reader consults another evidence holder.

\paragraph{Selection entropy and realized reward.}
If $p_{i,j}$ is agent $i$'s share of selections directed to ID $j$, its selection entropy is $-\sum_j p_{i,j}\log p_{i,j}$, which we average across the six selecting agents. This quantity ranges from zero for one repeated choice to $\log 6$ for uniform selection, while mean realized reward averages Equation~\ref{eq:reward} over the reads that occurred.

\paragraph{Pooled hindsight selection regret.}
For a lineage $\ell$, let $\widehat{\mu}_{\ell}(i,j)$ denote the mean reward observed when agent $i$ selected peer $j$, pooled across conditions within the same layer. Taking $b_{\ell}(i)=\max_j\widehat{\mu}_{\ell}(i,j)$ as the hindsight benchmark, we define regret for condition $a$ as

\begin{equation}
R^a_{\ell}=\sum_{(q,i,t)\in\mathcal{D}^{a,\ell}}
\left[b_{\ell}(i)-\widehat{\mu}_{\ell}\left(i,j^a_{q,i,t}\right)\right].
\label{eq:regret}
\end{equation}

This metric compares each choice with the peer that has the highest pooled observed mean after the run, using peer values estimated from the histories each policy produces. Lower regret indicates greater concentration on peers with high pooled rewards, which we assess alongside the realized reward and final accuracy of those choices. We make these in-sample comparisons within each model and layer, using a shared question set, condition pool, and hindsight benchmark.

\paragraph{Interaction diagnostics.}
To examine the step from consultation to revision, we compare the revised answer after each external read with the reader's and selected peer's answers from round $t-1$. We measure the selected answer's $F_1$, whether it scores above the reader's prior answer, and the reader's subsequent change in $F_1$. When the prior answers differ after normalization, we also measure \emph{answer agreement}: whether the revision matches the peer's prior answer. Recovery and damage track transitions to and from token-perfect $F_1=1$, which can differ from exact match. These exploratory diagnostics summarize the interactions produced by each policy, with eligible denominators reported explicitly and all comparisons aggregated within lineages.

\section{Experimental setup}
\label{sec:experimental_setup}

\subsection{Models, replication, and comparison conditions}
\label{sec:arms}

The stable-role experiment evaluates Qwen3-32B \citep{qwen3} and Gemma-4-31B-it \citep{gemma4} over 12 seeded lineages per model, with all six agents in a run sharing the same deployment and prompts. Every model, lineage, and condition receives the same ordered set of 80 questions: the first 80 usable entries in the HotpotQA distractor validation split after filtering for hard questions with at least 10 passages. Layer~1 uses the first 40 questions and Layer~2 the next 40, with the lineage seed determining passage assignments and selector behavior.

\paragraph{Layer 1: selector check.}
Layer~1 compares five policies on the first 40 questions. Alongside the LinUCB policy \texttt{mace}, we evaluate \texttt{ucb1}, which uses empirical reward and a count-based UCB bonus with coefficient $\sqrt{2}$ without the nine-feature context, and \texttt{pre\_defined}, which follows a fixed cycle in which each agent selects the next ID, producing zero selection entropy. The \texttt{in\_context} policy asks the language model whose response to read after showing each candidate's selection count, mean reward, and nine-feature context; invalid outputs fall back to the fixed cycle. Finally, \texttt{random} samples uniformly from all six IDs. The state learned by \texttt{mace} in each lineage initializes the memory conditions in Layer~2.

\paragraph{Layer 2: memory transfer.}
Layer~2 evaluates memory transfer on 40 disjoint questions using four conditions. Both \texttt{mace\_warm} and \texttt{mace\_shuffled} reuse the lineage's Layer~1 \texttt{mace} state: warm preserves the original peer IDs, and shuffled permutes them as described in Section~\ref{sec:peer_memory}. The comparison conditions are \texttt{mace\_cold}, which initializes new MACE state, and \texttt{random}, which samples uniformly and independently of learned state.

The shuffle moves every non-self record to a different ID while allowing transfers between the two evidence holders. Across the 12 lineages, 47 of 144 records learned about evidence holders (32.6\%) remain assigned to evidence-holder IDs at initialization, so this control tests a partial disruption of identity and role alignment. Appendix~\ref{app:shuffle} describes the retained alignment, and Appendix~\ref{app:reproducibility} gives implementation details, decoding settings, and response statistics.

\subsection{Evidence-role reassignment}
\label{sec:role_reassignment}

We test evidence-role reassignment in an additional study on Qwen3.5-9B \citep{qwen35}, served by Together AI with advertised FP8 quantization. Following the same fixed 40-question learning and 40-question evaluation blocks, the study uses six agents and three synchronous revisions. Each of 12 lineages learns fresh MACE state from empty initialization, with all priors completed before evaluation begins.

Evaluation crosses cold or warm initialization with stable or changed evidence-holder roles. Stable conditions retain the original passage assignment, while changed conditions swap both evidence holders' passages with two seeded former distractor holders so that old and current evidence-holder IDs are disjoint. The resulting assignment stays fixed over the 40 evaluation questions. Each warm condition loads a separate copy of the lineage's prior under the original IDs, and each cold condition starts empty; all conditions continue the reference-answer updates in Equation~\ref{eq:reward}. The intervention changes both readers' and peers' information ownership, while agents receive neither role labels nor peer passages.

The primary endpoint, fixed before inspecting this study's outcomes, is the paired final-$F_1$ interaction,
$(\mathrm{warm}-\mathrm{cold})_{\mathrm{changed}}-(\mathrm{warm}-\mathrm{cold})_{\mathrm{stable}}$.
Each condition contributes an average over 240 final agent responses per lineage. We report the mean interaction with a two-sided 95\% $t_{11}$ interval, conditional on the fixed questions and assuming independent, approximately normal lineage interactions. The remaining analyses---within-regime contrasts, reward, consultation rates, and five-question windows---are exploratory. Appendix~\ref{app:provider_study} gives the decoding and evaluation settings.

\subsection{Statistical analysis of the stable-role experiment}

We use the lineage as the unit of replication and pair all Layer~2 contrasts within lineages. The primary comparisons are warm minus cold for regret and final $F_1$, evaluated separately for each model; warm minus shuffled and all other metrics are exploratory. We report mean paired differences, two-sided 95\% $t_{11}$ intervals under an assumption of independent, approximately normal paired differences, and the number of lineages, out of 12, that favor warm. For regret, we express the mean difference and its interval as percentages of the model's mean cold regret across the 12 lineages.

The intervals are marginal and unadjusted for multiple comparisons, describing variation across the 12 seeds and passage assignments on the fixed question set. Separate API requests may contribute additional generation variability. Although the interaction diagnostics summarize many reads, these reads are selected by their respective policies and the sample size for inference remains 12 lineages.

Across all stable-role conditions, at least 99.98\% of responses yielded a valid answer, and all selector replies were parseable. Fewer than 0.4\% of responses reached the output limit in each model and layer (Appendix~\ref{app:reproducibility}).

\section{Results}
\label{sec:results}

We first report the stable-role experiment.

\subsection{Selector check}

In Layer~1, MACE has the lowest pooled hindsight regret on both models (Table~\ref{tab:layer1}, Appendix~\ref{app:selector_check}). UCB1 has higher exact match on Qwen3-32B and higher $F_1$, exact match, and realized reward on Gemma-4-31B-it, illustrating the complementary information provided by selection and answer measures. We therefore track both consultation and final answers when evaluating memory transfer in Layer~2.

\subsection{Persistent state changes consultation}

Warm memory lowers regret relative to cold on both models, with a paired difference of $-7.4\%$ of cold regret on Qwen3-32B (95\% CI $[-12.8,-2.1]$; 10 of 12 lineages favor warm) and $-6.0\%$ on Gemma-4-31B-it (CI $[-11.5,-0.5]$; 11 of 12 favor warm). Selection also becomes more concentrated: warm initialization has lower entropy in 11 of 12 lineages for each model. These results show that saved LinUCB state carries experience from the first task block into later choices.

\begin{figure}[!htbp]
    \centering
    \includegraphics[width=0.92\textwidth]{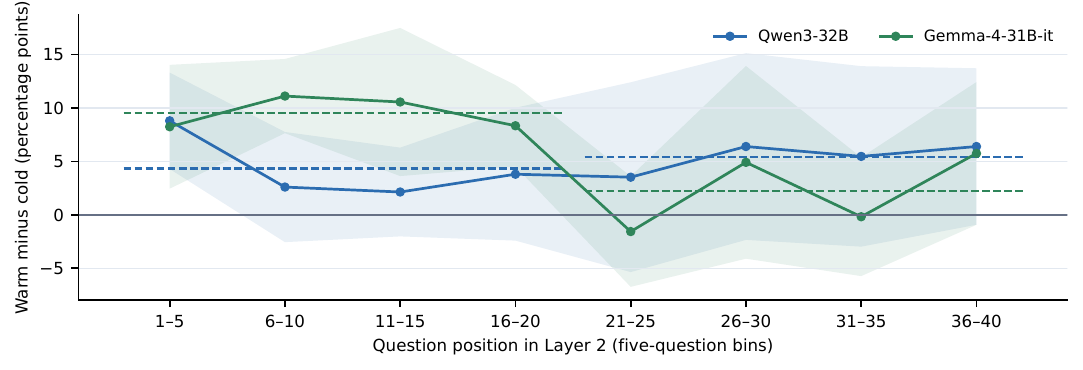}
    \caption{Warm minus cold difference in the probability of selecting an evidence-holder ID across Layer~2, including self-selection. Each point summarizes a five-question bin across 12 lineages, with shaded bands showing marginal, exploratory 95\% paired $t_{11}$ intervals and dashed lines showing half-run means. Qwen3-32B retains a 4--5 percentage point advantage, while the Gemma-4-31B-it advantage falls from 9.6 points to 2.2 points.}
    \label{fig:selection_trajectory}
\end{figure}

The evidence-holder read trajectories show how these choices evolve over the evaluation block (Figure~\ref{fig:selection_trajectory}). Counting all selections of evidence-holder IDs, including self-selection, warm Qwen3-32B agents exceed cold agents by 4.3 percentage points in the first half and 5.4 points in the second. For Gemma-4-31B-it, the corresponding gap narrows from 9.6 to 2.2 points.

Warm initialization also increases consultation of other evidence holders across the full run. Measured as a share of all choices, external evidence-holder reads rise by 4.88 percentage points on Qwen3-32B (95\% CI $[+0.77,+9.00]$; 8 of 12 lineages favor warm) and 4.09 points on Gemma-4-31B-it (CI $[+1.80,+6.37]$; 10 of 12 favor warm). Round-0 answers confirm that evidence holders begin with more useful information than distractor holders (Appendix~\ref{app:reproducibility}), giving these consultation changes a direct interpretation in terms of information access.

\FloatBarrier
\subsection{Final accuracy and realized reward}

The Layer~2 means and paired contrasts in Tables~\ref{tab:layer2} and~\ref{tab:contrasts} leave the effect of warm initialization on final accuracy uncertain. With one $F_1$ point equal to 0.01 on the reported scale, the warm-minus-cold estimate is $-1.26$ points on Qwen3-32B (95\% interval $[-3.15,+0.62]$; 4 of 12 lineages favor warm) and $+0.23$ points on Gemma-4-31B-it (interval $[-0.57,+1.04]$; 7 of 12 favor warm). Both intervals include zero, even as consultation shifts consistently toward evidence holders.

Realized reward follows a similar pattern on Qwen3-32B, where the warm-minus-cold difference is $-0.0053$, with an exploratory 95\% interval of $[-0.0157,+0.0050]$. Together with the accuracy estimates, this result shows how a measurable shift in consultation can coexist with uncertainty in task outcomes.

The shuffled-state comparison further shows a regret reduction on Qwen3-32B: warm minus shuffled is $-7.4\%$, with a CI of $[-12.3,-2.5]$. Gemma's regret contrast and both models' $F_1$ contrasts have intervals that include zero. These contrasts evaluate the partial disruption of role alignment described in Section~\ref{sec:arms}.

\begin{table}[!htbp]
\centering
\caption{Layer~2 means over 12 lineages. Regret totals are unnormalized and should be compared only within a model and layer. Entropy is descriptive. Bold marks the lowest pooled hindsight regret and the highest $F_1$, EM, and realized reward within each model.}
\label{tab:layer2}
\footnotesize
\setlength{\tabcolsep}{4pt}
\begin{tabular}{lrrrrr}
\toprule
condition & regret & $F_1$ & EM & reward & entropy \\
\midrule
\multicolumn{6}{l}{\textit{Qwen3-32B}} \\
\texttt{mace\_cold}     & 53.3062 & \textbf{0.3488} & \textbf{0.2868} & \textbf{0.1816} & 1.6203 \\
\texttt{mace\_warm}     & \textbf{49.3522} & 0.3362 & 0.2740 & 0.1762 & 1.5551 \\
\texttt{mace\_shuffled} & 53.2964 & 0.3359 & 0.2819 & 0.1763 & 1.5888 \\
\texttt{random}         & 73.5289 & 0.3164 & 0.2556 & 0.1611 & 1.7715 \\
\midrule
\multicolumn{6}{l}{\textit{Gemma-4-31B-it}} \\
\texttt{mace\_cold}     & 27.0813 & 0.1489 & \textbf{0.1226} & 0.0715 & 1.7181 \\
\texttt{mace\_warm}     & \textbf{25.4646} & \textbf{0.1513} & 0.1212 & \textbf{0.0736} & 1.6741 \\
\texttt{mace\_shuffled} & 26.7294 & 0.1441 & 0.1149 & 0.0702 & 1.6886 \\
\texttt{random}         & 32.3649 & 0.1382 & 0.1128 & 0.0668 & 1.7715 \\
\bottomrule
\end{tabular}
\end{table}

\begin{table}[!htbp]
\centering
\caption{Paired Layer~2 contrasts. Regret is expressed as a percentage of the same model's cold regret. Intervals are two-sided 95\% $t_{11}$ intervals. The last column counts lineages in which warm has lower regret or higher $F_1$. Warm minus cold is the primary contrast; warm minus shuffled is exploratory.}
\label{tab:contrasts}
\footnotesize
\setlength{\tabcolsep}{5pt}
\begin{tabular}{llrrr}
\toprule
metric and model & contrast & $\Delta$ & 95\% CI & warm favored \\
\midrule
\multicolumn{5}{l}{\textit{Regret, percentage of cold}} \\
Qwen3-32B & warm $-$ cold & $-7.4$ & $[-12.8,-2.1]$ & 10/12 \\
Qwen3-32B & warm $-$ shuffled & $-7.4$ & $[-12.3,-2.5]$ & 10/12 \\
Gemma-4-31B-it & warm $-$ cold & $-6.0$ & $[-11.5,-0.5]$ & 11/12 \\
Gemma-4-31B-it & warm $-$ shuffled & $-4.7$ & $[-11.2,+1.8]$ & 8/12 \\
\midrule
\multicolumn{5}{l}{\textit{Final token $F_1$}} \\
Qwen3-32B & warm $-$ cold & $-0.0126$ & $[-0.0315,+0.0062]$ & 4/12 \\
Qwen3-32B & warm $-$ shuffled & $+0.0003$ & $[-0.0181,+0.0187]$ & 5/12 \\
Gemma-4-31B-it & warm $-$ cold & $+0.0023$ & $[-0.0057,+0.0104]$ & 7/12 \\
Gemma-4-31B-it & warm $-$ shuffled & $+0.0072$ & $[-0.0012,+0.0156]$ & 8/12 \\
\bottomrule
\end{tabular}
\end{table}

\subsection{What happens after an external read?}
\label{sec:interaction}

The interaction traces connect the answers agents consult to the revisions that follow (Table~\ref{tab:interaction_diagnostics}). On Qwen3-32B, external reads that offer an answer scoring above the reader's prior answer increase from 18.69\% under cold initialization to 20.92\% under warm initialization, a paired difference of $+2.24$ percentage points with an exploratory interval of $[+0.40,+4.08]$. Because this opportunity measure compares the two prior answers, it reflects the quality of both reader and peer. The mean consulted-answer $F_1$ difference also has a positive estimate, with an interval that includes zero.

Among Qwen reads where reader and peer initially disagree, the revision matches the peer's answer in 27.36\% of eligible cold reads and 25.57\% of warm reads, giving a paired difference of $-1.79$ percentage points with an interval of $[-3.58,+0.01]$. Such agreement can accompany either a useful correction or adoption of an incorrect answer, so we also measure the change in the reader's score. Mean $F_1$ improvement per external read is 0.0239 under cold initialization and 0.0214 under warm initialization, with an uncertain paired difference. On Gemma, intervals include zero for changes in better-answer opportunities, answer agreement, and per-read improvement, and the mean revision change is slightly negative in both conditions.

% Generated from trace_diagnostics.csv by analyze_traces.py.
\begin{table}[!htbp]
\centering
\caption{Exploratory external-read diagnostics: means of 12 lineage estimates and paired warm-minus-cold 95\% $t_{11}$ intervals. Opportunity means the peer's prior $F_1$ exceeds the reader's. Agreement means the revision equals the peer's prior answer, conditional on different normalized prior answers. Eligible counts are in Appendix~\ref{app:interaction}. Rates use percentages, rate contrasts use percentage points, and $F_1$ uses the 0--1 scale.}
\label{tab:interaction_diagnostics}
\footnotesize
\setlength{\tabcolsep}{4pt}
\begin{tabular}{lrrrr}
\toprule
Diagnostic & Cold & Warm & $\Delta$ & 95\% CI \\
\midrule
\multicolumn{5}{l}{\textit{Qwen3-32B}} \\
Consulted prior $F_1$ & 0.3775 & 0.3910 & +0.0135 & [-0.0139, +0.0409] \\
Better-answer opportunity (\%) & 18.69 & 20.92 & +2.24 & [+0.40, +4.08] \\
Answer agreement (\%) & 27.36 & 25.57 & -1.79 & [-3.58, +0.01] \\
Revision $\Delta F_1$ & 0.0239 & 0.0214 & -0.0026 & [-0.0084, +0.0033] \\
\midrule
\multicolumn{5}{l}{\textit{Gemma-4-31B-it}} \\
Consulted prior $F_1$ & 0.1682 & 0.1746 & +0.0064 & [-0.0089, +0.0217] \\
Better-answer opportunity (\%) & 15.18 & 14.87 & -0.31 & [-1.50, +0.88] \\
Answer agreement (\%) & 15.74 & 15.67 & -0.07 & [-1.24, +1.10] \\
Revision $\Delta F_1$ & -0.0027 & -0.0022 & +0.0006 & [-0.0025, +0.0036] \\
\bottomrule
\end{tabular}
\end{table}

Appendix~\ref{app:interaction} complements these averages with token-perfect recovery and damage rates and examples of useful and harmful agreement. Taken together, the diagnostics allow us to follow consultation and revision as separate steps in the collaboration process.

\FloatBarrier
\subsection{Persistence after evidence reassignment}
\label{sec:reassignment_results}

The primary final-$F_1$ interaction in the Qwen3.5-9B study is $-1.38$ points, with a 95\% $t_{11}$ interval of $[-3.50,+0.73]$, leaving the effect of reassignment on warm initialization uncertain (Table~\ref{tab:provider-role-interaction}). Within regimes, the warm-minus-cold estimate is $-1.50$ points under changed roles, with an interval of $[-2.79,-0.20]$, and $-0.11$ points under stable roles, with an interval of $[-1.92,+1.70]$. These within-regime contrasts are exploratory and unadjusted for multiplicity.

\begin{table}[!htbp]
\centering
\small
\caption{Fresh Qwen3.5-9B study: final token $F_1$ means and paired differences across 12 lineages. Intervals use $t_{11}$, conditional on the fixed questions. The changed-minus-stable interaction is the primary target; within-regime contrasts are exploratory. Reassignment changes both readers' and peers' information ownership.}
\label{tab:provider-role-interaction}
\begin{tabular}{lrrl}
\toprule
Evidence roles & Cold & Warm & Warm--cold [95\% CI] \\
\midrule
Stable & 0.247 & 0.246 & -0.001 [-0.019, +0.017] \\
Changed & 0.253 & 0.238 & -0.015 [-0.028, -0.002] \\
\midrule
\multicolumn{3}{l}{Changed--stable interaction} & -0.014 [-0.035, +0.007] \\
\bottomrule
\end{tabular}
\end{table}

Using all choices as the denominator, warm initialization increases external current-evidence consultations by 3.45 percentage points under stable roles, with an exploratory interval of $[+0.75,+6.15]$. Under changed roles, the corresponding contrast is $-1.83$ points, with an interval of $[-6.02,+2.36]$, while warm agents self-select 5.00 points more often than cold agents, with an interval of $[+3.18,+6.82]$. Appendix~\ref{app:provider_study} extends these analyses with reward, both external-read denominators, all eight planned windows, and response validity.

\FloatBarrier
\section{Discussion}

Our results show that remembering peers changes whom agents consult: under stable evidence access, warm initialization lowers pooled hindsight regret and increases consultation of other evidence holders on both models. This consistency indicates that experience from earlier tasks carries forward into later choices. Assessing final answers alongside those choices gives a fuller account of memory's value, since the final-$F_1$ differences remain uncertain in this workflow.

The interaction traces help connect the consultation results to the answer-only communication protocol, in which each agent holds one passage and receives the selected peer's extracted answer. Warm Qwen agents more often encounter peer answers that score above their own prior answers, while changes in answer agreement and revision gain remain uncertain. This motivates controlled experiments that vary the information peers share and examine how readers use it.

The reassignment study broadens this evaluation to changing evidence access, using freshly learned memory and paired stable controls to examine what happens when different agents hold useful information. Under reassignment, warm agents self-select more often than cold agents, while the primary interaction leaves uncertain how reassignment changes the accuracy effect. Together with the shuffled-state comparison, this experiment provides a way to assess peer memory as the relationship between agent identity and information access changes.

\section{Limitations}
\label{sec:limitations}

The experiments use six agents that share a model and prompts, exchange answers, and learn from reference-answer feedback after every revision. The reassignment study introduces a different, smaller model and provider and changes each reader's own evidence access along with peer roles. Evaluating heterogeneous agents, gradual or repeated changes, richer messages, and rewards available without reference answers would establish how broadly the findings extend beyond these conditions.

The observed accuracy, especially on Gemma, reflects the answer-only workflow and may change with stronger reasoning or communication protocols. HotpotQA answers may also appear in pretraining, allowing both passage-based reasoning and memorized knowledge to contribute. Because each study uses the same fixed 80 questions and 12 lineages, its intervals describe variation on those questions; newly sampled questions and additional benchmarks would broaden the evidence.

The interaction diagnostics are observational, with message content, the reader's prior answer, and generation behavior all potentially affecting a revision. Hindsight regret likewise depends on the interactions each policy selects and cannot estimate how an unchosen peer would have helped in the same situation. Interpretation also depends on generation settings: the stable-role output cap is unknown, and the reassignment study uses a 192-token cap reached by 3.2\% of responses. Longer outputs and changes to hosted model versions may affect performance.

\section{Broader impacts}
\label{sec:broader_impacts}

Persistent peer records can preserve useful experience as well as mistaken beliefs, making it important to communicate where a record came from, which peer it describes, and when it should be reevaluated. Deployment also requires attention to privacy and data governance as interaction histories accumulate. Future evaluations should test these practices while tracking answer quality as information access changes.

\section{Conclusion}

PeerMem shows how experience with peers carries forward into later collaboration: reusing MACE's selector state lowers pooled hindsight regret and increases consultation of other evidence holders when roles stay fixed, while the effect on final accuracy remains uncertain. Interaction traces connect these choices to answer revisions, and the reassignment study extends the evaluation to changing evidence access. Together, the findings support evaluating peer memory through the full sequence of consultation, revision, and final task performance.

\clearpage
\begin{ack}
We thank the Algoverse AI Research program for its mentorship. This work received no funding. The authors declare no relevant competing interests.
\end{ack}

{\small
\bibliographystyle{plainnat}
\bibliography{references}
\par}

\clearpage
\appendix

\section{Layer 1 selector check}
\label{app:selector_check}
Layer~1 compares selectors on the first 40 questions, before persistence is introduced. These descriptive means characterize the selector used in the memory-transfer experiment. Regret totals use a shared hindsight reference within each layer and model.

\begin{table}[!htbp]
\centering
\caption{Layer~1 means over 12 lineages. Bold marks the lowest pooled hindsight regret and the highest $F_1$, EM, and realized reward within each model. Entropy is descriptive.}
\label{tab:layer1}
\footnotesize
\setlength{\tabcolsep}{4pt}
\begin{tabular}{lrrrrr}
\toprule
condition & regret & $F_1$ & EM & reward & entropy \\
\midrule
\multicolumn{6}{l}{\textit{Qwen3-32B}} \\
\texttt{mace}          & \textbf{41.6825} & \textbf{0.2932} & 0.2090 & \textbf{0.1506} & 1.6603 \\
\texttt{ucb1}          & 49.3733 & 0.2890 & \textbf{0.2122} & 0.1442 & 1.7706 \\
\texttt{pre\_defined} & 50.6789 & 0.2749 & 0.2014 & 0.1387 & 0.0000 \\
\texttt{in\_context}  & 54.8287 & 0.2520 & 0.1809 & 0.1275 & 0.4720 \\
\texttt{random}        & 54.3643 & 0.2733 & 0.1976 & 0.1398 & 1.7715 \\
\midrule
\multicolumn{6}{l}{\textit{Gemma-4-31B-it}} \\
\texttt{mace}          & \textbf{22.7451} & 0.1384 & 0.0972 & 0.0676 & 1.7260 \\
\texttt{ucb1}          & 25.4555 & \textbf{0.1402} & \textbf{0.0986} & \textbf{0.0678} & 1.7827 \\
\texttt{pre\_defined} & 26.2840 & 0.1281 & 0.0927 & 0.0626 & 0.0000 \\
\texttt{in\_context}  & 23.4727 & 0.1314 & 0.0951 & 0.0649 & 0.0257 \\
\texttt{random}        & 27.1468 & 0.1270 & 0.0899 & 0.0629 & 1.7715 \\
\bottomrule
\end{tabular}
\end{table}

\section{Additional stable-role contrast visualization}

\begin{figure}[!htbp]
    \centering
    \includegraphics[width=0.90\textwidth]{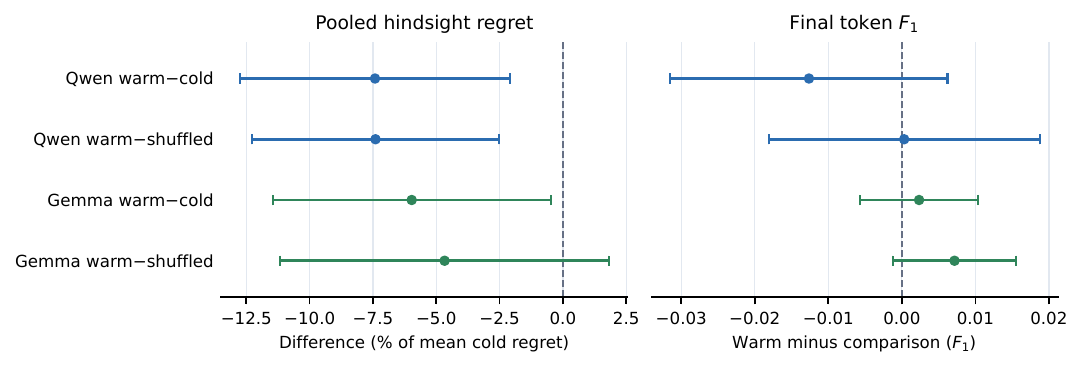}
    \caption{Paired differences with 95\% $t_{11}$ intervals. Negative regret and positive final $F_1$ favor warm. Three regret intervals exclude zero, while all four final-$F_1$ intervals include zero.}
    \label{fig:paired_contrasts}
\end{figure}

\FloatBarrier
\section{Implementation and reproducibility details}
\label{app:reproducibility}

\paragraph{Implementation and code availability.}
We implement the MACE selector described in its July 2026 paper \citep{mace2026}. We will release the experiment code, configurations, prompt templates, and analysis scripts in a public GitHub repository, including the question-selection and seeded role-assignment procedures. Our selection analysis uses pooled hindsight regret as defined in Section~\ref{sec:metrics}; MACE's October revision reports an empirical realized-reward gap against Random \citep{mace2026v2}.

\paragraph{Context features.}
For selecting agent $i$, candidate peer $j$, and revision round $t$, the MACE context is

\begin{equation}
x_{i,j,t}=\left[d^{(1)}_{i,j},d^{(2)}_{i,j},d^{(3)}_{i,j},c^{(1)}_j,c^{(2)}_j,c^{(3)}_j,\widehat{p}_{i,j},\tau_t,1\right]^\top.
\end{equation}

The diversity feature $d^{(n)}_{i,j}$ is one minus the Jaccard similarity between the agents' $n$-gram sets, and $c^{(n)}_j$ gives peer $j$'s normalized share of total response-graph divergence at order $n$. Historical reward enters as $\widehat{p}_{i,j}=(\mu_{i,j}+0.5)/1.5$, where $\mu_{i,j}$ is the pair's mean reward; the remaining entries are the normalized revision round $\tau_t=t/T$ and a bias. Before scoring, we scale the vector as $x\leftarrow x/\max(1,\lVert x\rVert_2)$. LinUCB initializes $A_{ij}=I_9$ and $b_{ij}=0$, uses exploration coefficient $\alpha=1$, and updates the selected ordered pair after each reward. For $N$ agents and $d$ features, these matrices and vectors require $O(N^2d^2)$ numeric storage, and selection requires no additional language-model call.

\paragraph{Prompt information flow.}
At revision round 0, each agent receives the question and its assigned passage and returns a concise response inside an \texttt{answer} tag. Subsequent rounds add the agent's previous answer and the selected peer's extracted answer to this input. Extraction uses the last closed answer tag, falling back to the trailing text of an unclosed final tag or the last nonempty line when no tag is present. For \texttt{in\_context} selection, the model receives each candidate's selection count, mean reward, and nine-feature context before choosing whose answer to read. Appendix~\ref{app:prompts} provides the templates used for these calls.

\paragraph{Models and decoding.}
We serve Qwen3-32B and Gemma-4-31B-it through Azure Foundry with temperature 0 and \texttt{top\_p} left unspecified, keeping the deployment and prompts fixed across conditions within each model. Qwen receives the \texttt{/no\_think} prompt suffix, and Gemma receives no suffix. The client defaults to 192 output tokens, although the effective limit used in these runs was not recorded. Qwen runs on a dedicated two-H100 deployment; Gemma's serving hardware is unspecified.

\paragraph{Experiment size and response length.}
Each model is evaluated on 12 lineages with six agents and 40 questions per layer, with each agent producing an initial response and three revisions per question. The five Layer~1 conditions require 66,240 model calls per model, including in-context selection, and the four Layer~2 conditions require 46,080, for a total of 224,640 calls across both models. Depending on layer and condition, extracted answers average 2.91--3.14 whitespace-delimited words for Qwen and 4.08--4.48 for Gemma.

Responses reached the output limit in 0.11\% and 0.23\% of Qwen calls in Layers~1 and 2, respectively, and in 0.39\% and 0.28\% of Gemma calls. These responses remain in the analysis and pass through the same parser as all other responses.

\paragraph{Question selection and evidence-holder roles.}
The data loader scans the HotpotQA distractor validation split in its stored order and takes the first 80 hard questions with at least 10 passages, preserving the same question IDs and order across models, lineages, and conditions. Each lineage seed determines which two IDs receive evidence, and this assignment is shared across the lineage's stable-role conditions.

\paragraph{Evidence-role validation.}
Round-0 responses confirm that agents assigned evidence begin with substantially more useful information. Across Layer~2 conditions, their $F_1$ ranges from 0.520 to 0.532 on Qwen3-32B and from 0.367 to 0.369 on Gemma-4-31B-it, compared with distractor-holder ranges of 0.151 to 0.157 and 0.051 to 0.052, respectively. These role labels come directly from the passage assignment.

\paragraph{Existing assets and licenses.}
HotpotQA is distributed under CC BY-SA 4.0 \citep{hotpotqa2018}.\footnote{\url{https://hotpotqa.github.io/}} The public Qwen3-32B and Gemma-4-31B-it releases carry Apache 2.0 licenses and are cited through their technical reports \citep{qwen3,gemma4}.\footnote{\url{https://huggingface.co/Qwen/Qwen3-32B}; \url{https://huggingface.co/google/gemma-4-31B-it}} We do not train or release a new model or dataset.

\section{Evidence-role reassignment study}
\label{app:provider_study}

\paragraph{Model and decoding.}
Together AI serves \texttt{Qwen/Qwen3.5-9B} with advertised FP8 quantization and a 262,144-token total context limit. We set temperature to 0, \texttt{max\_tokens} to 192, and \texttt{enable\_thinking=false} in \texttt{chat\_template\_kwargs}, with no suffix added to the prompts. Answering and revision prompts, extraction, self-selection, and MACE updates follow Section~\ref{sec:method}, and peer selection requires no language-model calls. Every response used in the analysis reports the requested model ID and zero reasoning tokens, although the provider did not supply an immutable served-weight identifier or hardware details. The public model release carries an Apache 2.0 license.\footnote{\url{https://huggingface.co/Qwen/Qwen3.5-9B}}

\paragraph{Protocol and reproducibility.}
The learning phase uses 12 lineages, 40 questions, six agents, and four generations per agent, yielding 11,520 answer calls; the four evaluation conditions add 46,080 calls, bringing the total to 57,600. Passage assignments, reassignment, and scheduling use fixed seeds, with cold/warm order and stable/changed order balanced across lineages. Conditions run sequentially, with six concurrent agent requests in each synchronous round, and each warm condition receives its own copy of the completed prior. Although API generations may vary even at temperature 0, the configurations, question order, responses, and selector states support replay of scoring and policy updates.

\paragraph{Consultation denominators.}
Old-holder and current-holder choice rates divide by all revisions and include self-selection. For external current-evidence consultations, we report rates using both all external reads and all choices as denominators, with the latter matching the denominator in the stable-role experiment. Each evaluation condition contains 720 revisions per lineage, and each five-question window contains 30 final answers and 90 revisions; external-read counts depend on the policy's choices. Across these analyses, the unit of replication remains the 12 lineages.

\paragraph{Response validity and length limits.}
All 57,600 raw responses used in the analysis are nonempty and none is content-filtered, but the parser returns an empty answer for 25 generations, which remain in the analysis with $F_1=0$. Invalid-extraction counts for learning, stable cold, stable warm, changed cold, and changed warm are 2, 9, 8, 5, and 1, respectively, out of 11,520 responses per condition. Among the 11,520 evaluation final answers, three are invalid---two in changed cold and one in stable cold---with none in either warm cell. Of all responses, 1,838 (3.19\%) reach the length limit, and all receive normal scoring.

Table~\ref{tab:provider-output-limit} reports length-limit counts across all generation rounds, showing 19 fewer capped responses in changed warm than in changed cold. Because round-specific and final-answer cap counts are unavailable, the influence of truncation on final scores remains uncertain.

\begin{table}[!htbp]
\centering
\small
\caption{Qwen3.5-9B responses that reached the output limit across 12 lineages and all four generation rounds. Each condition contains 11,520 responses, and all responses remain in the analysis.}
\label{tab:provider-output-limit}
\begin{tabular}{lrrr}
\toprule
Cell & Length-limited & Generations & Rate (\%) \\
\midrule
Learning & 340 & 11,520 & 2.95 \\
Stable cold & 340 & 11,520 & 2.95 \\
Stable warm & 373 & 11,520 & 3.24 \\
Changed cold & 402 & 11,520 & 3.49 \\
Changed warm & 383 & 11,520 & 3.32 \\
\midrule
Total & 1,838 & 57,600 & 3.19 \\
\bottomrule
\end{tabular}
\end{table}

\begin{figure}[p]
\centering
\includegraphics[width=\textwidth]{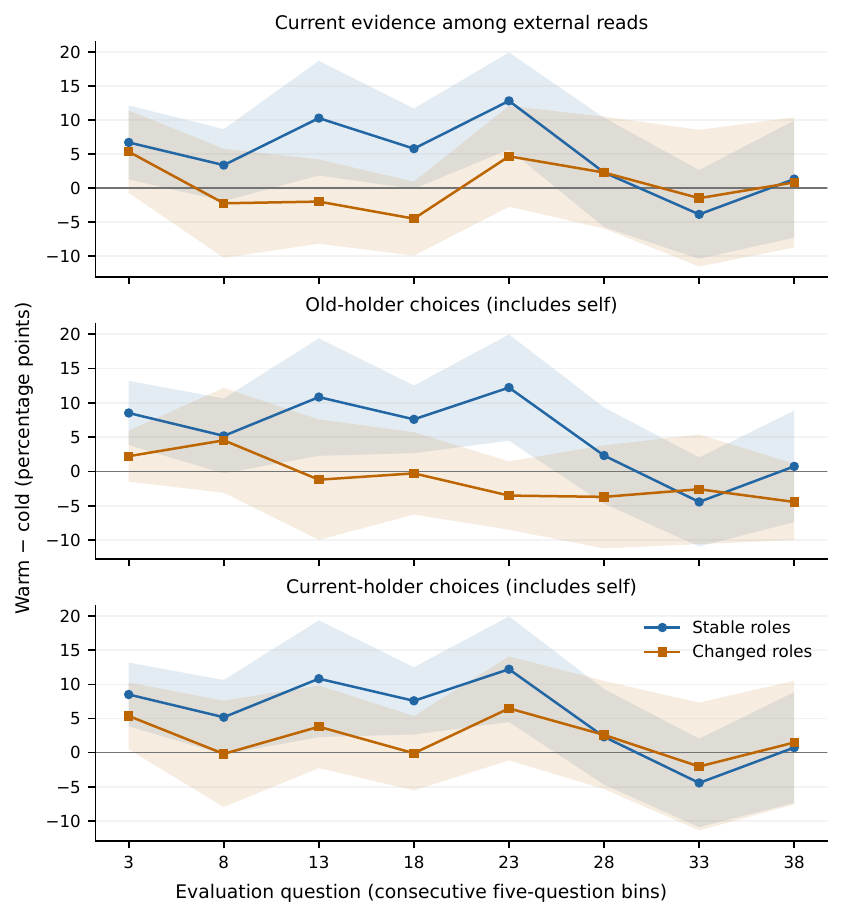}
\caption{Exploratory Qwen3.5-9B warm-minus-cold consultation differences over all eight five-question bins, with marginal paired 95\% $t_{11}$ bands. The top panel divides external current-evidence reads by external reads; the lower panels divide old/current-holder selections by all choices and include self-selection. Stable old and current holders coincide; changed old and current holders are disjoint.}
\label{fig:provider_selection_trajectory}
\end{figure}

\paragraph{Exploratory consultation and reward.}
Table~\ref{tab:provider-exploratory} reports consultation and reward contrasts with marginal, unadjusted $t_{11}$ intervals. The warm-minus-cold difference in old-holder choices under changed roles is $-1.12$ percentage points, with an interval of $[-4.03,+1.78]$, and the whole-run reward interaction is $-0.0065$, with an interval of $[-0.0185,+0.0055]$. Figure~\ref{fig:provider_selection_trajectory} traces the consultation contrasts across all eight planned five-question windows.

\begin{table}[!htbp]
\centering
\scriptsize
\caption{Exploratory whole-run paired contrasts with 95\% Student intervals. Rate differences are percentage points. External current-E among external reads divides by external reads; its companion divides by all choices. Inclusive holder and self rates also divide by all choices. Retained CSV analysis records contain every numerator, denominator, and eligible lineage.}
\label{tab:provider-exploratory}
\begin{tabular}{lp{.26\linewidth}p{.26\linewidth}}
\toprule
Metric & Stable warm--cold & Changed warm--cold \\
\midrule
Reward & -0.000 [-0.010, +0.010] & -0.006 [-0.015, +0.002] \\
External current-E (external reads) & +4.587 [+1.941, +7.232] & +0.424 [-4.616, +5.463] \\
External current-E (all choices) & +3.449 [+0.745, +6.153] & -1.829 [-6.018, +2.361] \\
Old-holder choices (includes self) & +5.370 [+2.790, +7.950] & -1.123 [-4.028, +1.782] \\
Current-holder choices (includes self) & +5.370 [+2.790, +7.950] & +2.176 [-2.727, +7.079] \\
Self choices & +1.400 [-0.145, +2.946] & +5.000 [+3.176, +6.824] \\
\bottomrule
\end{tabular}
\end{table}

\begin{figure}[!htbp]
\centering
\includegraphics[width=0.78\textwidth]{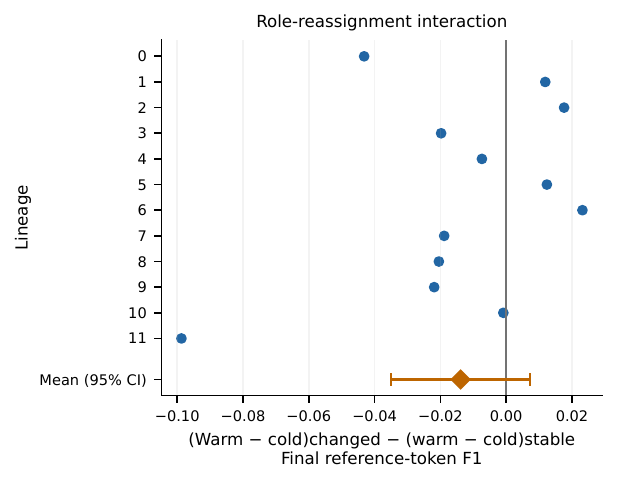}
\caption{Qwen3.5-9B final-$F_1$ interaction by lineage. Each point is the changed-minus-stable difference in the warm-minus-cold contrast. The summary gives the mean and 95\% $t_{11}$ interval across 12 lineages, conditional on the fixed evaluation questions.}
\label{fig:provider_f1_interaction}
\end{figure}

\FloatBarrier

\section{Interaction diagnostics and illustrative traces}
\label{app:interaction}

For each Layer~2 revision, we compare the reader's new answer with its own and the selected peer's answers from the preceding round, using evidence roles from the seeded passage assignment. Scores are computed against the reference answer, and rewards follow Equation~\ref{eq:reward}.

% Generated from trace_diagnostics.csv by analyze_traces.py.
\begin{table}[t]
\centering
\caption{External-read token-perfect transitions. Recovery conditions on previous reader $F_1<1$ and requires revised $F_1=1$; damage conditions on previous $F_1=1$ and requires revised $F_1<1$. $N_c,N_w$ give eligible cold/warm reads across all lineages. Values are lineage-mean percentages; contrasts and paired 95\% $t_{11}$ intervals use percentage points. Transitions are defined using token $F_1$.}
\label{tab:token_perfect_transitions}
\footnotesize
\setlength{\tabcolsep}{4pt}
\begin{tabular}{llrrrrr}
\toprule
Model & Transition & $N_c$ & $N_w$ & Cold & Warm & $\Delta$ [95\% CI] \\
\midrule
Qwen & Recovery & 5884 & 5936 & 7.01 & 6.76 & -0.25 [-1.38, +0.88] \\
Qwen & Damage & 2152 & 2071 & 10.54 & 11.58 & +1.04 [-1.64, +3.72] \\
Gemma & Recovery & 6799 & 6723 & 1.50 & 1.38 & -0.11 [-0.62, +0.40] \\
Gemma & Damage & 875 & 841 & 9.37 & 9.55 & +0.18 [-1.90, +2.25] \\
\bottomrule
\end{tabular}
\end{table}

% Generated from trace_diagnostics.csv by analyze_traces.py.
\begin{table}[t]
\centering
\caption{External reads by receiver and selected-peer passage role. $N_c,N_w$ count eligible reads for revision change; $F_1$ change is on the 0--1 scale. Estimates first average within each lineage. Contrasts pair the 12 lineage estimates with 95\% $t_{11}$ intervals. Role labels follow the seeded passage assignment.}
\label{tab:role_diagnostics}
\footnotesize
\setlength{\tabcolsep}{3pt}
\begin{tabular}{lrrrrr}
\toprule
Receiver $\to$ peer & $N_c$/$N_w$ & Cold $\Delta F_1$ & Warm $\Delta F_1$ & Difference & 95\% CI \\
\midrule
\multicolumn{6}{l}{\textit{Qwen3-32B}} \\
Evidence $\to$ evidence & 939/1139 & 0.0356 & 0.0251 & -0.0105 & [-0.0261, +0.0052] \\
Evidence $\to$ distractor & 1702/1502 & -0.0190 & -0.0219 & -0.0030 & [-0.0167, +0.0108] \\
Distractor $\to$ evidence & 3056/3278 & 0.0550 & 0.0582 & +0.0033 & [-0.0099, +0.0165] \\
Distractor $\to$ distractor & 2339/2088 & 0.0054 & -0.0072 & -0.0126 & [-0.0282, +0.0030] \\
\midrule
\multicolumn{6}{l}{\textit{Gemma-4-31B-it}} \\
Evidence $\to$ evidence & 686/773 & 0.0015 & 0.0007 & -0.0007 & [-0.0098, +0.0084] \\
Evidence $\to$ distractor & 1774/1531 & -0.0316 & -0.0293 & +0.0023 & [-0.0055, +0.0101] \\
Distractor $\to$ evidence & 2302/2568 & 0.0226 & 0.0213 & -0.0013 & [-0.0081, +0.0055] \\
Distractor $\to$ distractor & 2912/2692 & -0.0065 & -0.0108 & -0.0043 & [-0.0099, +0.0012] \\
\bottomrule
\end{tabular}
\end{table}

For consulted-answer quality, better-answer opportunities, and revision change, the external-read denominators are 8,036 cold and 8,007 warm reads on Qwen, and 7,674 cold and 7,564 warm reads on Gemma. Agreement excludes identical normalized prior answers, leaving 4,608 cold and 4,836 warm reads on Qwen, and 4,334 cold and 4,172 warm reads on Gemma. Token-perfect recovery is conditional on prior $F_1<1$; damage is conditional on prior $F_1=1$. All comparisons use 12 lineage-level averages.

\paragraph{Illustrative transitions.}
The examples below show the first recovery and first damage event in each model's warm condition, ordered by lineage, question, round, and reader. Both examples per model occur in lineage~0; aggregate rates appear in the preceding tables.

\begin{table}[!htbp]
\centering
\caption{Deterministically selected warm-condition examples. The first arrow links the reader's previous answer to its revision; the peer column gives the selected previous-round answer. E and D indicate actual evidence and distractor roles. Each recovery changes $F_1$ from 0 to 1; each damage example changes it from 1 to 0.}
\label{tab:trace_examples}
\footnotesize
\setlength{\tabcolsep}{4pt}
\begin{tabular}{p{1.1cm}>{\raggedright\arraybackslash}p{5.1cm}>{\raggedright\arraybackslash}p{3.3cm}>{\raggedright\arraybackslash}p{2.3cm}}
\toprule
Model & Question and reader transition & Peer answer and roles & Reference \\
\midrule
Qwen & What coastal area borders Dirleton's medieval fortress?\newline North Sea $\rightarrow$ Yellowcraig & Yellowcraig\newline reader D, peer E & Yellowcraig \\
\addlinespace
Qwen & Where is Sachin Warrier's former software employer headquartered?\newline Mumbai $\rightarrow$ Unknown & Unknown\newline reader E, peer D & Mumbai \\
\addlinespace
Gemma & Which Midwest track hosts a 500-mile race each May?\newline I do not have enough information to answer this question. $\rightarrow$ Indianapolis Motor Speedway & Indianapolis Motor Speedway\newline reader D, peer E & Indianapolis Motor Speedway \\
\addlinespace
Gemma & Are Ferocactus and Silene both types of plant?\newline Yes $\rightarrow$ Insufficient information & Insufficient information\newline reader D, peer E & yes \\
\bottomrule
\end{tabular}
\end{table}

The questions are shortened for presentation, and the examples illustrate how agreement can accompany either a correction or an error depending on the answer a peer supplies.

\FloatBarrier
\clearpage
\section{Prompt templates}
\label{app:prompts}

The stable-role experiments use the following templates, with one system message and one user message in each model call. Qwen appends \texttt{/no\_think} to each user message, while Gemma uses no suffix. For each call, the braced fields are filled with the corresponding question, passage, responses, or candidate information.

\paragraph{Discussion system message.}
\begin{Prompt}
You are one of several agents answering a question. Each agent has been given a different piece of context, and no single piece is guaranteed to be relevant.
\end{Prompt}

\paragraph{Initial user message.}
\begin{Prompt}
Context:
{context}
Question: {question}
First, briefly reason over the provided context. Then, place your final answer between <answer> and </answer> tags. Keep the answer concise - a word, phrase, or short sentence. Example: <answer>Paris</answer>
\end{Prompt}

\paragraph{Revision user message.}
\begin{Prompt}
Context:
{context}
Question: {question}
YOUR PREVIOUS RESPONSE:
{own_previous}
RESPONSE FROM {peer_name}:
{peer_previous}
INSTRUCTIONS:
Review your previous response and the peer's response carefully. You may agree, disagree, or selectively incorporate the peer's reasoning. Then provide your updated, complete response to the QUESTION.
First, briefly reason over the provided context. Then, place your final answer between <answer> and </answer> tags. Keep the answer concise - a word, phrase, or short sentence. Example: <answer>Paris</answer>
\end{Prompt}

The context contains the title and text of the agent's assigned passage, and both previous-response fields contain extracted answers. Thus, although the revision instructions invite agents to incorporate a peer's reasoning, the information supplied excludes the peer's generated reasoning and source passage.

\paragraph{In-context selector system message.}
\begin{Prompt}
You are selecting which peer's response to read.
\end{Prompt}

\paragraph{In-context selector user message.}
The user message starts with the following lines:
\begin{Prompt}
QUESTION: {question}

HISTORY:
\end{Prompt}
One row follows for each of the six candidates, including the selecting agent itself:
\begin{Prompt}
Agent {k} [chosen {n_k} times, mean reward: {mu_k} | features: [Diversity (unigram): {x_1}, Diversity (bigram): {x_2}, Diversity (trigram): {x_3}, Centrality (unigram): {x_4}, Centrality (bigram): {x_5}, Centrality (trigram): {x_6}, Hist. Reward: {x_7}, Norm. Round: {x_8}, Bias: {x_9}]]
\end{Prompt}
Each row summarizes the agent's previous selections of that candidate through a selection count and mean reward. The nine features describe the candidate's answer diversity relative to the selecting agent and its centrality in the response graph, each measured with unigrams, bigrams, and trigrams. They also include the pair's historical reward, the revision round, and a bias term. We rescale the historical-reward feature and normalize the feature vector, while leaving the separately displayed mean reward unscaled. After the candidate rows, the message ends with:
\begin{Prompt}
Based on the historical information, whose full response would you like to read to help refine your answer? You may also choose your own.
Reply with only the agent number (1 to 6).
\end{Prompt}
This LLM selector is used in Layer~1, where the selected candidate's extracted answer is passed to the revision.

\section{Shuffled-state control}
\label{app:shuffle}

For each selecting agent, the shuffled condition keeps the self record in place and assigns each of the other five records to a different peer ID, using a permutation determined by the lineage seed and selecting-agent ID. All learned statistics move together, after which online updates continue under the new assignment.

The shuffle changes peer identities while preserving some alignment with evidence access: across 12 lineages, 23 of 120 non-self records learned about evidence holders move to another evidence-holder ID. Together with the 24 fixed self records, this leaves 47 of 144 evidence-origin records (32.6\%) assigned to evidence holders, defining the degree of role disruption tested by the control.

A follow-up test could isolate non-self attribution by resetting self records in both conditions and comparing correctly indexed state with a role-aware permutation that moves every evidence-origin record to a distractor ID and gives every evidence-holder ID a distractor-origin record. With two evidence holders and four distractors, the remaining transfers would occur between distractors.

\clearpage
\section*{NeurIPS Paper Checklist}

\begin{enumerate}

\item {\bf Claims}
    \item[] Question: Do the main claims made in the abstract and introduction accurately reflect the paper's contributions and scope?
    \item[] Answer: \answerYes{} % Replace by \answerYes{}, \answerNo{}, or \answerNA{}.
    \item[] Justification: The Abstract and Introduction distinguish the stable-role study from the fresh-provider reassignment study, describe consultation separately from answer quality, and retain uncertainty in the primary interaction.

\item {\bf Limitations}
    \item[] Question: Does the paper discuss the limitations of the work performed by the authors?
    \item[] Answer: \answerYes{} % Replace by \answerYes{}, \answerNo{}, or \answerNA{}.
    \item[] Justification: Section~\ref{sec:limitations} discusses the stable and reassigned evidence settings, answer-only exchange, low absolute accuracy, possible benchmark memorization, supervised reference-answer feedback, limited replication, observational interaction diagnostics, exploratory analyses, and the limits of hindsight regret.

\Needspace{4\baselineskip}
\item {\bf Theory assumptions and proofs}
    \item[] Question: For each theoretical result, does the paper provide the full set of assumptions and a complete (and correct) proof?
    \item[] Answer: \answerNA{} % Replace by \answerYes{}, \answerNo{}, or \answerNA{}.
    \item[] Justification: The paper contains no theoretical results, theorems, or proof claims.

    \item {\bf Experimental result reproducibility}
    \item[] Question: Does the paper fully disclose all the information needed to reproduce the main experimental results of the paper to the extent that it affects the main claims and/or conclusions of the paper (regardless of whether the code and data are provided or not)?
    \item[] Answer: \answerNo{} % Replace by \answerYes{}, \answerNo{}, or \answerNA{}.
    \item[] Justification: The paper describes the experimental protocol, prompt templates, and analyses, although the effective output-token limit for the stable-role runs and the exact hosted model weights are unavailable.

\item {\bf Open access to data and code}
    \item[] Question: Does the paper provide open access to the data and code, with sufficient instructions to faithfully reproduce the main experimental results, as described in supplemental material?
    \item[] Answer: \answerNo{} % Replace by \answerYes{}, \answerNo{}, or \answerNA{}.
    \item[] Justification: HotpotQA is publicly available and cited. We plan to release the experiment code, configurations, and analysis scripts in a public GitHub repository; public code access is pending that release.

\item {\bf Experimental setting/details}
    \item[] Question: Does the paper specify all the training and test details (e.g., data splits, hyperparameters, how they were chosen, type of optimizer) necessary to understand the results?
    \item[] Answer: \answerYes{} % Replace by \answerYes{}, \answerNo{}, or \answerNA{}.
    \item[] Justification: The experiments involve no model training. Sections~\ref{sec:method} and~\ref{sec:experimental_setup}, together with Appendices~\ref{app:reproducibility} and~\ref{app:provider_study}, specify the data split, agents, conditions, parameters, prompt information flow, model identifiers, decoding, and analysis needed to understand the results.

\item {\bf Experiment statistical significance}
    \item[] Question: Does the paper report error bars suitably and correctly defined or other appropriate information about the statistical significance of the experiments?
    \item[] Answer: \answerYes{} % Replace by \answerYes{}, \answerNo{}, or \answerNA{}.
    \item[] Justification: Section~\ref{sec:experimental_setup} defines the stable-role and role-reassignment lineage-level analyses and marginal two-sided 95\% $t_{11}$ intervals, and Section~\ref{sec:results} reports the corresponding estimates and replication counts.
    \par\Needspace{3\baselineskip}

\item {\bf Experiments compute resources}
    \item[] Question: For each experiment, does the paper provide sufficient information on the computer resources (type of compute workers, memory, time of execution) needed to reproduce the experiments?
    \item[] Answer: \answerNo{} % Replace by \answerYes{}, \answerNo{}, or \answerNA{}.
    \item[] Justification: Appendix~\ref{app:reproducibility} reports experiment size and the two-H100 Qwen deployment. Execution time and the serving hardware for Gemma and Together are unspecified.
    
\item {\bf Code of ethics}
    \item[] Question: Does the research conducted in the paper conform, in every respect, with the NeurIPS Code of Ethics \url{https://neurips.cc/public/EthicsGuidelines}?
    \item[] Answer: \answerYes{} % Replace by \answerYes{}, \answerNo{}, or \answerNA{}.
    \item[] Justification: Sections~\ref{sec:method} and~\ref{sec:limitations}, together with the Broader Impacts section, document the public-benchmark study, its limitations, and its risks. No human participants or private interaction histories were collected.

\Needspace{4\baselineskip}
\item {\bf Broader impacts}
    \item[] Question: Does the paper discuss both potential positive societal impacts and negative societal impacts of the work performed?
    \item[] Answer: \answerYes{} % Replace by \answerYes{}, \answerNo{}, or \answerNA{}.
    \item[] Justification: Section~\ref{sec:broader_impacts} discusses mistaken peer records, record provenance and reevaluation, and privacy and data governance when interaction histories persist. Future work can evaluate these deployment practices.
    
\item {\bf Safeguards}
    \item[] Question: Does the paper describe safeguards that have been put in place for responsible release of data or models that have a high risk for misuse (e.g., pre-trained language models, image generators, or scraped datasets)?
    \item[] Answer: \answerNA{} % Replace by \answerYes{}, \answerNo{}, or \answerNA{}.
    \item[] Justification: The work releases no new dataset or model and does not introduce an artifact with a high risk of misuse.

\item {\bf Licenses for existing assets}
    \item[] Question: Are the creators or original owners of assets (e.g., code, data, models), used in the paper, properly credited and are the license and terms of use explicitly mentioned and properly respected?
    \item[] Answer: \answerYes{} % Replace by \answerYes{}, \answerNo{}, or \answerNA{}.
    \item[] Justification: The paper cites HotpotQA and all three model releases. Appendices~\ref{app:reproducibility} and~\ref{app:provider_study} state the CC BY-SA 4.0 dataset license and public Apache 2.0 model licenses.

\item {\bf New assets}
    \item[] Question: Are new assets introduced in the paper well documented and is the documentation provided alongside the assets?
    \item[] Answer: \answerNA{} % Replace by \answerYes{}, \answerNo{}, or \answerNA{}.
    \item[] Justification: A public code release is planned, but no new assets are publicly released with the current manuscript.

\item {\bf Crowdsourcing and research with human subjects}
    \item[] Question: For crowdsourcing experiments and research with human subjects, does the paper include the full text of instructions given to participants and screenshots, if applicable, as well as details about compensation (if any)? 
    \item[] Answer: \answerNA{} % Replace by \answerYes{}, \answerNo{}, or \answerNA{}.
    \item[] Justification: The paper involves neither crowdsourcing nor research with human subjects.

\Needspace{8\baselineskip}
\item {\bf Institutional review board (IRB) approvals or equivalent for research with human subjects}
    \item[] Question: Does the paper describe potential risks incurred by study participants, whether such risks were disclosed to the subjects, and whether Institutional Review Board (IRB) approvals (or an equivalent approval/review based on the requirements of your country or institution) were obtained?
    \item[] Answer: \answerNA{} % Replace by \answerYes{}, \answerNo{}, or \answerNA{}.
    \item[] Justification: The paper involves no research with human subjects, so IRB or equivalent review is not applicable.

\item {\bf Declaration of LLM usage}
    \item[] Question: Does the paper describe the usage of LLMs if it is an important, original, or non-standard component of the core methods in this research? Note that if the LLM is used only for writing, editing, or formatting purposes and does \emph{not} impact the core methodology, scientific rigor, or originality of the research, declaration is not required.
    %this research? 
    \item[] Answer: \answerYes{} % Replace by \answerYes{}, \answerNo{}, or \answerNA{}.
    \item[] Justification: Sections~\ref{sec:method} and~\ref{sec:experimental_setup}, with Appendices~\ref{app:reproducibility} and~\ref{app:provider_study}, describe Qwen3-32B, Gemma-4-31B-it and Qwen3.5-9B as answering agents and the LLM-based in-context selector.

\end{enumerate}

\end{document}